# Quantitative Analysis of Demand Response Using Thermostatically Controlled Loads

Praveen Dhanasekar, *Student Member, IEEE*, Cunzhi Zhao, *Student Member, IEEE* and Xingpeng Li, *Member, IEEE*

{pdhanasekar@uh.edu; czhao20@uh.edu; xli83@central.uh.edu}

Department of Electrical and Computer Engineering

University of Houston,

Houston, Texas, USA.

*Abstract*—The flexible power consumption feature of thermostatically controlled loads (TCLs) such as heating, ventilation, and air-conditioning (HVAC) systems makes them attractive targets for demand response (DR). TCLs possess a brief period where their power utilization can be altered without any significant impact on customer comfort level. This indicates TCLs are hidden potentials for providing ancillary services. This paper proposes a novel metric of demand response support time (DRST) for HVAC enabled demand response and a novel algorithm for the quantification of such HVAC-DR. The consumers' comfort will not be compromised with the proposed DRST-based HVAC-DR. Case studies demonstrate its benefits in terms of cost saving in microgrid day-ahead scheduling and reduction of forced load shedding during a grid-microgrid tie-line outage event. This illustrates the reserve potential benefits and the increase of microgrid reliability when DRST-based HVAC-DR is considered.

## Nomenclature

| | |
|---|---|
| $A_T$ | Ambient temperature (°F or °C) |
| $Q$ | Heat rate for HVAC unit (Btu/°F) |
| $W$ | Heat exchange factor for house objects (Btu/°F) |
| $A_{HC}$ | Air heat capacity (Btu/°F or J/°C) |
| $M_{HC}$ | Mass heat capacity (Btu/°F or J/°C) |
| $A_R$ | Thermal resistance of air (°F/Btu*hr or °C/ W) |
| $M_R$ | Thermal resistance of mass (°F/Btu*hr or °C/ W) |
| $H_T$ | Air temperature inside the house (°F or °C) |
| $\dot{H}_T$ | Differentiation of $H_T$ with respect to time |
| $T_{SET}$ | Set/desired temperature (°F or °C) |
| $M_T$ | Mass temperature inside the house (°F or °C) |
| $F_{cost}$ | Cost function of the day-ahead scheduling ($) |
| $C_{gt}$ | Cost of microturbine $g$ for period $t$ ($) |
| $C_t^{Grid}$ | Cost of grid power for period $t$ ($/kWh) |
| $P_t^{Grid}$ | Trading power with the grid in period $t$ (kW) |
| $P_{TieLine}$ | Tie-line power limit exchange with the grid (kW) |
| $P_{st}$ | Solar power produced in period $t$ (kW) |
| $P_{gt}$ | Generated power of microturbine $g$ in period $t$ (kW) |
| $L_{HVAC(t)}$ | HVAC load consumption in period $t$ (kW) |
| $L_{Non-HVAC(t)}$ | Non-HVAC load consumption in period $t$ (kW) |
| $L_{Shed(t)}$ | Load shed in period $t$ (kW) |
| $L_t^{DR}$ | The amount of DR deployment after the occurrence of tie-line outage event (kW) |
| $P_{et}^C$ | Charging power of BESS $e$ in period $t$ (kW) |
| $P_{et}^D$ | Discharging power of BESS $e$ in period $t$ (kW) |
| $P_e^{CMax}$ | Maximum charging power of BESS $e$ (kW) |
| $P_e^{DMax}$ | Maximum discharging power of BESS $e$ (kW) |
| $P_h^{HVAC}$ | Power capacity of HVAC $h$ (kW) |
| $T^R$ | Pre-specified contingency response time |
| $R_g^{up}$ | Maximum ramp up rate of microturbine $g$ (kW) |
| $R_g^{down}$ | Maximum ramp down rate of microturbine $g$ (kW) |
| $RS_{gt}$ | Reserve provided by generator $g$ in period $t$ |
| $RS_{TOTAL(t)}$ | Total reserve available for microgrid in period $t$ |
| $RS_{et}$ | Reserve provided by BESS $e$ in period $t$ |
| $RS_t^{HVAC}$ | Reserve provided by HVAC in period $t$ |
| $DR_t^{HVAC}$ | Available demand response provided by HVAC in period $t$ |
| $E_{e0}$ | Initial energy stored at the beginning of first-time interval |
| $E_{et}$ | Energy stored in BESS $e$ in period $t$ (kWh) |
| $\eta_e^C$ | Charging efficiency of BESS $e$ (%) |
| $\eta_e^D$ | Discharging efficiency of BESS $e$ (%) |
| $E_e^{max}$ | Maximum energy limit of BESS $e$ (%) |
| $E_e^{Min}$ | Minimum energy limit of BESS $e$ (%) |
| $U_{et}^C$ | Charging status indicator variable of BESS $e$ in period $t$ |
| $U_{et}^D$ | Discharging status indicator variable of BESS $e$ in period $t$ |
| $U_{gt}$ | Status indicator of microturbine $g$ in period $t$ |
| $V_{gt}$ | Startup indicator of microturbine $g$ at time $t$ |

## I. Introduction

Due the worldwide energy polices of decreasing the proportion of carbon-intensive power generation, the renewable energy sources (RES) are growing fast and have surpassed coal, a conventional generation resource, since 2019. The stochastic and intermittent generation caused by high RES penetration will increase the difficulty to operate the power system [1]. Energy storage systems (ESS) are one of the solutions for mitigating uncertain power fluctuation. However, the high cost of ESS limits the installed capacity [2]. As a result, the grid may require more ancillary services to maintain system reliability.

Demand response is one of the ancillary services provided on the consumer end via reduction and shift of electricity consumption. Thermostatically controlled loads (TCLs) such as heating, ventilation, and air-conditioning (HVAC) system and water heater are ones of the most power consumption residential loads in the United States. A potential reserve can be provided to the grid by aggregating the TCLs [3]-[4].

Abundant research has proved that TCLs can be aggregated to provide demand response to the grid theoretically. A priority-stack-based control framework is proposed in [5] to manage the TCLs. Hierarchical control mechanism is designed in [6] to make the TCLs as manageable resources. Regulation services can be provided by aggregated HVAC control with smart thermostats in [7]. [8] presents a demand management algorithm by coordinating the battery energy storage system with HVAC. However, they all consider TCLs for direct load control which may affect users' comfort level. The proposed strategy in [9] can obtain the rates of temperature increase and decrease and improve the control of the TCLs, which makes it easier to determine the flexible power consumption of TCLs. An improved TCLs population model is described in [10] to provide a control algorithm considering the lockout time of HVAC; however, the heat transfer model is not taken into consideration. For those control strategies mentioned in [5]-[10], they either develop the strategies for direct load control or improve the deterministic algorithm for flexible TCLs; none of them provides a solution of potential demand response for each TCL without discomforting the consumer end.

To address the gap mentioned above, a novel metric demand response support time (DRST) and a DRST-based HVAC-DR quantification (DRST-HDRQ) algorithm are proposed in this paper. The proposed metric DRST quantifies demand response potential provided by HVAC by indicating how long an HVAC will be available for demand response at a given time, and thus, it can identify the set of HVACs that are qualified to provide DR without compromising consumers' comfort. The thermo dynamic model for the HVAC unit is implemented to determine the DRST for each HVAC in each time period. Then, the proposed DRST-HDRQ algorithm can precisely quantify the amount of demand response provided by qualified HVACs, which is the amount of load that can be deferred to the next scheduled interval without affecting customer comfort level. In addition, microgrid day-ahead scheduling and outage response are simulated in this paper to demonstrate the benefits of DRST-based HVAC-DR. It is worth mentioning that the proposed metric DRST and the HVAC-DR quantification algorithm can also be applied to other types of TCLs with modifications.

The rest of the paper is organized as follows. The HVAC system model, the proposed metric DRST and the proposed DRST-HDRQ algorithm are presented in Section *II*. The mathematical modeling of microgrid day-ahead scheduling considering DRST-based HVAC-DR is presented in Section *III*. The test microgrid case is described in Section *IV*. Section *V* analyzes the simulation results and demonstrates the benefits provided by HVAC enabled DR. Section *VI* concludes the paper.

## II. HVAC System Modeling

The modeling of TCL's heat transfer process is critical to relate it to the electricity consumption of the unit, which is presented below. The term *f(Q, W)* represents the function involving *Q* and *W* during the heat exchange process.

### A. Thermal Dynamics Model of an HVAC Unit

An equivalent thermal parameter (ETP) model for residential HVAC units is shown in Figure 1 [11]. This model is appropriate to be used for residential and small commercial buildings. A state space description of the ETP model [11] is as follows,

$$\dot{z} = K * z + L * x \quad (1)$$

$$\dot{z} = \begin{bmatrix} \dot{H}_T \\ \dot{M}_T \end{bmatrix} \quad (2)$$

$$x = 1 \quad (3)$$

$$K = \begin{bmatrix} -(\frac{1}{M_R * A_{HC}} + \frac{1}{A_R * M_{HC}}) & \frac{1}{M_R * A_{HC}} \\ \frac{1}{M_R * M_{HC}} & -\frac{1}{M_R * M_{HC}} \end{bmatrix} \quad (4)$$

$$L = \begin{bmatrix} \frac{A_T}{(A_R * A_{HC})} + \frac{Q}{A_{HC}} \\ 0 \end{bmatrix} \quad (5)$$

$$\dot{H}_T = \left(\frac{-2 * H_T}{R * C}\right) + \left(\frac{A_T}{R * C} + \frac{W}{C} - \frac{Q}{C}\right) \quad (6)$$

$$\dot{H}_T = \left(\frac{-2 * H_T}{R * C}\right) + \left(\frac{A_T}{R * C} + \frac{W}{C}\right) \quad (7)$$

where $\dot{z}$ represents the linear non-homogenous differential system with state vector; *K* represents the state matrix; and *L\*x* represents the time varying coefficient vector. The ETP model represented by (1)-(5) can be simplified by assuming the thermal resistance of air is equal to the thermal resistance of mass and assuming the thermal capacitance of air is equal to the thermal capacitance of mass. The equivalent thermal resistance and thermal capacitance are denoted as *R* and *C* respectively. The temperature calculation during the ON and OFF states of the HVAC are expressed in (6) and (7) respectively. The objects such as curtains and furniture inside the house radiate heat raising the temperature; this factor is accounted by the *W* term.

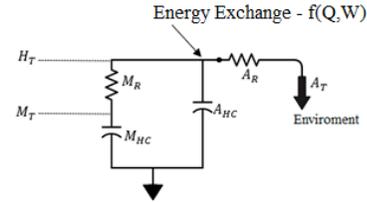

Figure 1. Equivalent thermal parameter model of a residential HVAC unit [11].

### B. DRST -based HVAC-DR Quantification

With the above two equations, the house's temperature is simulated for the whole day according to their temperature set points. The temperature dead-band is the maximum permissible temperature difference around the set temperature. The power levels of HVAC oscillate between the "On" state and "Off" state. The demand response support time is defined as the time that after an HVAC is turned off from "On" status at a given time point, the room temperature will take to reach the upper bound temperature of the HVAC setting if cooling or the lower bound temperature of the HVAC setting if heating.

Figure 2 illustrates the proposed metric DRST for HVAC at cooling mode. With a DR signal is received at $t_{DR}^{start}$, the HVAC can provide DR until it reaches the upper bound temperature at $t_{DR}^{end}$. DRST is the time period between $t_{DR}^{start}$ and $t_{DR}^{end}$. In Figure 2, $t_1, t_3, t_5$ are the extreme high points and $t_2, t_4$ are the extreme low points for the pre-set thermostat temperature respectively. Extreme low points are defined as the time points when the house temperature reaches the lower bound temperature of the HVAC setting while the extreme high points correspond to the upper bound temperature. The HVAC is at "on" state during time periods $(t_1, t_2) \& (t_3, t_4)$ and at "off" state during $(t_2, t_3) \&, (t_4, t_5)$.

A metric similar to DRST is the theoretically maximum DRST (TM-DRST). It gives us the idea about how long each HVAC can support when they receive DR signals at extreme low points (when cooling) or extreme high points (when heating).

When HVAC receives the DR signal right before those extreme low points such as $t_2^-$ or $t_4^-$ in Figure 2, the corresponding DRST is the TM-DRST. Note that like DRST, TM-DRST is also affected by several factors including the ambient temperature.

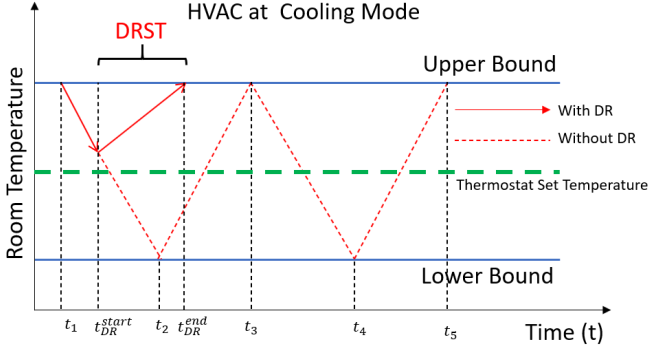

Figure 2. Illustration of the proposed metric DRST.

The temperature, DRST, power consumption under normal conditions and demand response provided by HVACs can be simulated and calculated by the following steps that are also illustrated in Figure 3. We consider a period of 24 hours with 10-minute resolution, which makes a total of 144 time intervals.

i. Thermal dynamic simulation for each HVAC for 24 hours.
ii. Set $DR_t^{HVAC} = 0$ for each time interval $t$ and $h=1$.
iii. Select HVAC $h$, and set $t=1$.
iv. Calculate DRST for this HVAC.
v. If $DRST > T^R$, HVAC $h$ is qualified to provide DR, updated $DR_t^{HVAC} = DR_t^{HVAC} + P_h^{HVAC}$; otherwise, this HVAC does not provide any DR for this time interval.
vi. If all time intervals checked, go to next step; otherwise, set $t=t+1$ and go to step $iv$.
vii. If all HVACs are checked, algorithm ends; otherwise, set $h=h+1$ and go to step $iii$.

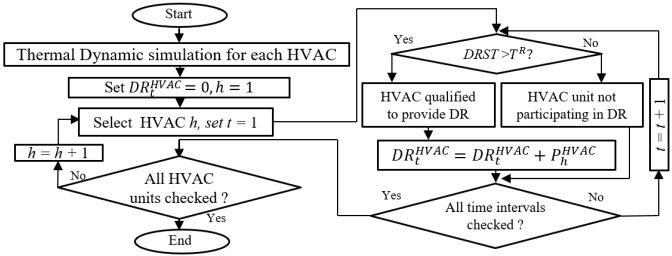

Figure 3. Flowchart of the proposed DRST-HDRQ algorithm.

## III. MATHEMATICAL MODEL

This section presents the day-ahead energy scheduling model for a networked microgrid. The microgrid normal day-ahead scheduling model consists of (8)-(32).

Its objective is to minimize the total cost of microgrid operations as shown in (8). Equation (9) describes the cost of distributed units in the microgrid. The power balance equation is shown in (10). Constraint (11) enforces the power limits of distributed units. The ramp up and ramp down rates are constrained by (12) and (13) respectively. The power exchanged with the bulk grid is limited in (14). Constraint (15) restricts the battery energy storage systems (BESS) to be either in charging mode or in discharging mode or stay idle. Constraints (16)-(17) limit the BESS charging and discharging power. The energy stored in BESS is calculated by (18)-(19). (20) assumes the ending energy level is equal to the initial energy level for BESS. The maximum and minimum energy levels of the BESS are ensured in (21). Equations (22)-(25) define the relationship between binary variables $U_{gt}$ and $V_{gt}$. Equation (26) defines the minimum microgrid reserve requirement to be 25% of the total demand. Equation (27) calculates the total reserve that is available for time interval $t$. The reserve provided by BESS is limited as shown in (28) and (29). Equation (30) limits the reserve provided by HVAC enabled demand response. $DR_t^{HVAC}$ can be obtained from the DRST-based HVAC-DR quantification algorithm as shown in Figure 3. Equations (31) and (32) limits the unit's power production in consideration with the reserve provided by the same unit.

Objective:
$$\min F_{cost} = \Sigma_t(\Sigma_g C_{gt} + P_t^{Grid} * C_t^{Grid}) \quad (8)$$

Constraints are as follows:
$$C_{gt} = c_g * U_{gt} + b_g * P_{gt} + a_g * P_{gt}^2 \quad (9)$$
$$P_t^{Grid} + P_{st} + \Sigma_g P_{gt} = L_{HVAC} + L_{Non-HVAC} + \Sigma_e(P_{et}^C - P_{et}^D) \; (\forall t) \quad (10)$$
$$P_g^{Max} U_{gt} \geq P_{gt} \geq P_g^{Min} U_{gt} \; (\forall g, t) \quad (11)$$
$$P_{g,t+1} - P_{gt} \leq R_g^{up} \; (\forall g, t) \quad (12)$$
$$P_{gt} - P_{g,t+1} \leq R_g^{down} \; (\forall g, t) \quad (13)$$
$$-P_{TieLine} \leq P_t^{Grid} \leq P_{TieLine} \; (\forall t) \quad (14)$$
$$U_{et}^C + U_{et}^D \leq 1 \; (\forall t, e) \quad (15)$$
$$0 \leq P_{et}^C \leq P_{et}^{CMax} * U_{et}^C \; (\forall t, e) \quad (16)$$
$$0 \leq P_{et}^D \leq P_{et}^{DMax} * U_{et}^D \; (\forall t, e) \quad (17)$$
$$E_{e1} = E_{e0} + (\eta_e^C * P_{e1}^C - P_{e1}^D/\eta_e^D)\Delta T \quad (18)$$
$$E_{et} = E_{e(t-1)} + (\eta_e^C * P_{et}^C - P_{et}^D/\eta_e^D)\Delta T \; (\forall e, t \geq 2) \quad (19)$$
$$E_{e0} = E_{eFinal} \quad (20)$$
$$E_e^{max} \geq E_{et} \geq E_e^{min} \; (\forall e, t) \quad (21)$$
$$V_{gt} \geq U_{gt} - U_{g(t-1)} \; (\forall g, t) \quad (22)$$
$$V_{g1} \geq U_{g1} \; (\forall g) \quad (23)$$
$$V_{g(t+1)} \leq 1 - U_{gt} \; (\forall g, t) \quad (24)$$
$$V_{gt} \leq U_{gt} \; (\forall g, t) \quad (25)$$
$$RS_{TOTAL(t)} \geq 25\% * (L_{HVAC(t)} + L_{Non-HVAC(t)}) \; (\forall t) \quad (26)$$
$$RS_{TOTAL(t)} = P_{TieLine} - P_t^{Grid} + RS_{gt} + RS_{et} + RS_t^{HVAC} \quad (27)$$
$$0 \leq RS_{et} \leq P_{et}^{DMax} - P_{et}^D + P_{et}^C \; (\forall e, t) \quad (28)$$
$$RS_{et} \leq \eta_e^D(E_e^{max} - E_{et})/\Delta T \; (\forall e, t) \quad (29)$$
$$RS_t^{HVAC} \leq DR_t^{HVAC} \; (\forall t) \quad (30)$$
$$P_g^{Min} * U_{gt} \leq P_{gt} + RS_{gt} \; (\forall g, t) \quad (31)$$
$$P_g^{Max} * U_{gt} \geq P_{gt} + RS_{gt} \; (\forall g, t) \quad (32)$$

When considering microgrid emergent operations under grid-microgrid tie-line disconnection event, the post-outage microgrid power balance constraint with DR deployment and forced load shed terms is shown in (33). Note that reserve requirements would no longer be enforced for emergent microgrid adjustment during the tie-line outage event.
$$P_{st} + \Sigma_g P_{gt} = L_{HVAC} + L_{Non-HVAC} - L_t^{DR} - L_{Shed(t)} + \Sigma_e(P_{et}^C - P_{et}^D) \; (\forall t) \quad (33)$$

## IV. TEST CASE DESCRIPTION

The microgrid used in this paper consists of solar power, two micro-turbines, and battery energy storage systems. The total solar capacity is 300 kW which is calculated based on the dataset obtained from Pecan Street Dataport [12]. There are two BESS systems with an energy capacity of 200 kWh and 150 kWh respectively. They share the same charging and discharging efficiencies that are 80% and 95% respectively. The microgrid contains 200 houses, of which the internal temperatures varying from 69°F to 78.8°F. The thermal resistance varies between 0.0654 °C/W and 0.0909 °C/W with an average of 0.0773 °C/W. The heat capacity varies from 3599.3 J/°C to 4500 J/°C with an average of 4074.9 J/°C. The HVAC's heat rate varies from 3000 J/°C to 2200 J/°C with an average of 2582.6 J/°C; and their real power ratings can be 2 kW, 2.21 kW, or 2.5 kW which are distributed in the ratio of 54%, 24%, 22% over all 200 houses. A temperature dead-band of 10°F is considered.

## V. CASE STUDIES

A day-ahead scheduling problem is solved for the microgrid system with 25% renewable energy described above. Test cases with and without HVAC-DR as reserve are considered in this paper and emergent operations of microgrid in the case of tie-line disconnection event are analyzed. The local generators in the microgrid are bound to provide the energy required to supply all the demand and maintain the reserve conditions set by the microgrid operator to accommodate the disconnection of main grid from the microgrid. The reserve requirement is set as 25% of the total demand. The microgrid is analyzed with different renewable energy penetration levels (30%, 40% and 50%).

### A. HVAC TM-DRST

The TM-DRST for each HVAC is represented in Figure 4. Table I describes the statistics of TM-DRST which depicts that the houses have different thermal characteristics which leads to different TM-DRST. The average TM-DRST is around 16 minutes over all 200 HVACs considered in this paper. The TM-DRST that an HVAC can achieve is 27 minutes while the lowest time duration that one of the HVACs can withstand without compromising consumer comfort is 10 minutes.

Table I Statistics about Microgrid HVACs' TM-DRST (minutes)

|  | Highest | Lowest | Average | Median | Standard Deviation |
|---|---|---|---|---|---|
| TM-DRST | 27 | 10 | 15.99 | 15 | 3.74 |

### B. HVAC enabled Demand Response

The demand response potential is calculated using the procedure explained in Figure 3. The emergency response time is set to be 10 minutes. This indicates that HVACs that have an DRST of more than 10 minutes (can be turned off from on status and be kept off for at least 10 minutes without compromising consumers' comfort) are qualified to provide demand response services to quickly mitigate microgrid power mismatch. For the microgrid test case used in this paper, more than 20% of HVAC units is available for most of the time for demand response for a 10-minute emergency response time as shown in Figure 5; particularly, for only 4 intervals or 2.8% of all intervals, the available HVAC units for DR is less than 20%.

### C. Microgrid Day-Ahead Scheduling

The microgrid day-ahead scheduling is conducted for two scenarios, where (i) the first scenario does not consider HVACs providing reserve, i.e. set $DR_t^{HVAC}$ to be zero in (30); and (ii) the second scenario incorporates the support of HVAC-DR to meet the reserve requirements. The inclusion of HVACs reduces the operating cost by 15% for the microgrid with 25% renewable energy which is evident from

Table II; for the cases with different renewable energy levels, it can still substantially reduce the cost. Figure 6 shows the power output of one of the two microturbines, which is similar to the other microturbine. The results from the day-ahead scheduling show that the sudden spikes in microturbine power generation were considerably reduced which is evident as shown in Figure 6. The load during the morning and afternoon hours were tackled with solar power and the power from the grid.

Table II Day-ahead scheduling results

| Renewable Energy Penetration Level | Scenario 1: without HVAC reserve | Scenario 2: HVAC as reserve |
|---|---|---|
| Base case (25%) | $1016.9 | $870.0 |
| 30% | $901.71 | $824.23 |
| 40% | $677.23 | $580.59 |
| 50% | $526.15 | $424.58 |

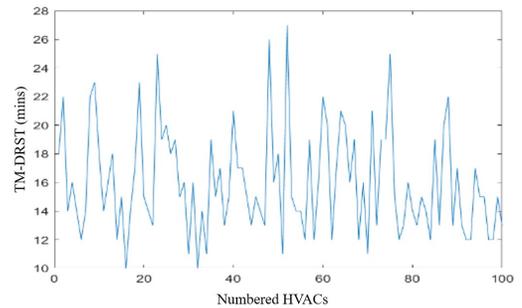

Figure 4. TM-DRST (minutes) for HVAC.

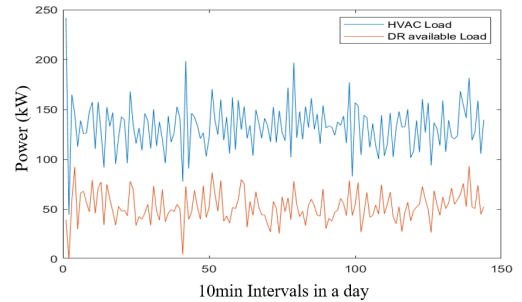

Figure 5. HVAC load and HVAC-DR potential.

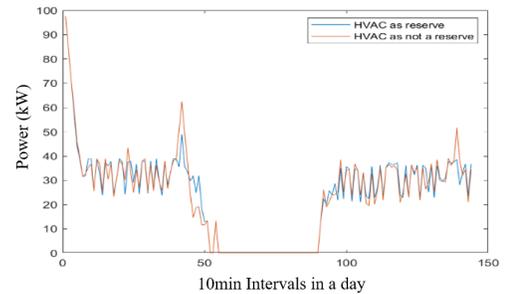

Figure 6. Power generation of one of the two micro-turbines.

### D. Tie-Line Outage Analysis

The microgrid utilizes maximum power from the bulk grid during period of minimal solar power. Moreover, a prolonged maximum tie-line usage is seen starting from hour 21 to hour 22 with minimal solar power and has one zero BESS power usage interval. Hence, this period is chosen for tie-line failure studies due to microgrid exploiting the maximum capacity of power grid. Multiple case scenarios with different renewable energy penetration levels are examined to understand load shed process during the tie-line disconnection period.

*a) Without consideration of HVAC-DR as reserve in microgrid day-ahead scheduling*

The pre-outage situation considered here is based on the solution obtained from scenario 1 as stated in sub-section *V.C*. The system is analyzed for two cases: the first case deploys HVAC-DR during load shedding process while the second case does not deploy HVAC-DR. From Table III, we can see that when using HVAC-DR as post-outage emergency control only (not serve as reserve providers in the microgrid normal day-ahead scheduling), it can still substantially reduce the amount of forced load shed under different penetration levels of renewable energy in the microgrid. For the microgrid base case with 25% renewable energy, the forced load shed under tie-line outage is reduced by 87.6% when deploying post-outage HVAC-DR as compared to not deploying post-outage HVAC-DR.

Table III Average forced load shed for scenario 1 without HVAC as reserve in microgrid day-ahead scheduling

| Renewable energy penetration level | Average forced load Shed | |
|---|---|---|
| | Without post-outage HVAC-DR deployment (kW) | With post-outage HVAC-DR deployment (kW) |
| Base case (25%) | 56.27 | 6.96 |
| 30% | 49.99 | 3.90 |
| 40% | 32.50 | 4.51 |
| 50% | 27.93 | 2.8729 |

Table IV Average forced load shed for scenario 2 with HVAC as reserve in microgrid day-ahead scheduling

| Renewable energy penetration level | Average forced load shed | |
|---|---|---|
| | Without post-outage HVAC-DR deployment (kW) | With post-outage HVAC-DR deployment (kW) |
| Base case (25%) | 58.03 | 6.96 |
| 30% | 53.89 | 5.13 |
| 40% | 39.16 | 7.45 |
| 50% | 54.11 | 6.66 |

*b) With the consideration of HVAC-DR as reserve in microgrid day-ahead scheduling*

The pre-outage situation considered here is based on the solution obtained from scenario 2 as stated in sub-section *V.C*. The system in the post-outage situation is analyzed for two cases that (1) deploys HVAC-DR and (2) does not deploy HVAC-DR during microgrid emergent operations under tie-line outage event. The average forced load shed over multiple outages under different renewable energy penetration levels are presented in Table IV for both cases. From Table IV, we can see that the sum of average forced load shed over four different renewable energy penetration levels are 26.2 kW and 205.19 kW for those two cases respectively. It is clear to conclude that under the tie-line failure event, the average forced load shed can be significantly reduced if HVAC enabled DR is deployed and it does not even compromise consumers' comfort.

Thus, it is evident from Table III and Table IV that in addition to meeting reserve requirement, HVACs can also be made advantageous for supporting emergent control and improving microgrid resilience.

## VI. CONCLUSION

To facilitate quantitative analysis of HVAC enabled DR, a new metric DRST is proposed in this paper. With this metric, a novel HVAC enabled DR quantification algorithm is proposed in this paper. In addition, this paper examines the benefits of utilizing DRST-based HVAC enabled DR and provides sufficient evidence on how HVACs can be used to provide microgrid ancillary services or demand response without compromising customers comfort level. Numerical simulations in Section *V* demonstrate that HVACs are excellent source of reserve potential as demand response providers. The simulations on microgrid day-ahead scheduling demonstrated 15% reduction in the operating cost when including HVAC-DR as reserve. When HVAC-DR is deployed as an emergent response action against bulk grid-microgrid tie-line outage, the amount of outage-induced forced load shed can be substantially reduced. This reduces the intensity of power continuity disruptions to appreciable limits. Moreover, the proposed strategy can be applied to other types of TCLs. Overall, this work demonstrates the capability of thermostatically controlled loads specifically HVACs to achieve cost saving, alleviate sudden generation spikes, and reduce forced load shed under bulk grid-microgrid tie-line disconnection event.